\documentclass[%
  reprint,
superscriptaddress,
 amsmath,
 longbibliography,
 aps,
 pra,
]{revtex4-2}
\usepackage{color}
\usepackage{mhchem}
\usepackage{graphicx}% Include figure files
\usepackage{dcolumn}% Align table columns on decimal point
\usepackage{bm}% bold math
\usepackage[colorlinks=true,citecolor=blue,urlcolor=blue,linkcolor=blue]{hyperref}% add hypertext capabilities
\usepackage[mathlines]{lineno}% Enable numbering of text and display math
\usepackage{siunitx}

\begin{document}
%===============================================
%%\preprint{}
\title{Ultrafast Relaxation Dynamics of Inner-Shell Vacancies in Hydrated Pyrrole}
%===============================================
\author{Kedong Wang}
\email{wangkd@htu.edu.cn}
\affiliation {\it School of Physics, Henan Normal University, Xinxiang 453007, People’s Republic of China}
\author{Bohui Wan}
\email{wanbohui@stu.htu.edu.cn}
\affiliation {\it School of Physics, Henan Normal University, Xinxiang 453007, People’s Republic of China}
\author{Cody L. Covington}
\email{covingtonc@apsu.edu}
\affiliation{ Department of Chemistry, Austin Peay State University, Clarksville, 
Tennessee 37044, United States}
\author{K\'alm\'an Varga}
\email{kalman.varga@vanderbilt.edu}
\affiliation{ Department of Physics and Astronomy, Vanderbilt University, Nashville, 
Tennessee 37235, United States}

\date{\today}
%===============================================
%===============================================
\begin{abstract}
%===============================================
We employ real-space, real-time time-dependent density functional theory (TDDFT) combined with Ehrenfest dynamics to investigate ultrafast intermolecular relaxation following inner-valence ionization in hydrated pyrrole. This time-dependent approach treats electronic and nuclear motions simultaneously, allowing the description of electronic excitation, charge transfer, ionization, and nuclear motion.
When the initial vacancy in the O 2s$^{-1}$ state is created on the water molecule, the system predominantly undergoes intermolecular Coulombic decay (ICD) and electron-transfer–mediated decay (ETMD), accompanied by pronounced charge transfer between pyrrole and water. In contrast, ionization of the pyrrole site for N 2s electron leads to both ICD and Auger decay channels. These results demonstrate that the decay dynamics are strongly governed by the initial vacancy location, offering microscopic insight into intermolecular energy-transfer mechanisms in hydrated molecular systems.

%==============================================
\end{abstract}
%==============================================
\maketitle
%==============================================
\section{Introduction}
%==============================================
Inner-valence ionization is a fundamental process in radiation chemistry and photophysics, particularly in aqueous environments relevant to biological systems \cite{skitnevskaya2023, gao2025}. It removes electrons from orbitals located between the core and outer-valence levels, producing excited cationic states that relax through ultrafast channels such as Auger decay, intermolecular Coulombic decay (ICD), and electron-transfer–mediated decay (ETMD) \cite{moddeman1971, jahnke2004, averbukh2006, zobeley2001}. In hydrated clusters, solvent interactions decisively influence these pathways by modulating charge redistribution and energy transfer dynamics \cite{skitnevskaya2023, Kumar2025}.

Auger decay, first observed in the 1920s, occurs when an inner-shell vacancy is filled by an electron from a higher orbital, and the released energy ejects a secondary electron. The resulting multiply charged molecular ions are highly unstable and typically fragment within tens of femtoseconds \cite{ITALA2011}. In biological contexts, such fragmentation may lead to severe local damage and compromise DNA integrity, thereby impacting radiation therapy outcomes \cite{boudaiffa2000}. For isolated molecules with inner-valence vacancies, Auger decay is often energetically forbidden; however, in the presence of neighboring species, the excess energy can be released via interatomic or intermolecular Coulombic decay.

The introduction of ICD mechanism by Cederbaum $et$ $al$. in 1997 substantially extended the understanding of electronic relaxation in weakly bound systems \cite{cederbaum1997, santra2001}. In ICD, the excess energy is transferred nonradiatively from an ionized site to a neighboring molecule, leading to the emission of a low-energy electron capable of inducing secondary damage \cite{gao2025}. ICD has been experimentally confirmed in rare-gas clusters, water clusters, and, more recently, in liquid water, where it competes with proton transfer and other nonadiabatic relaxation processes \cite{Marburger2003, ohrwall2004, Zhang2025, Ren2023}. Moreover, Mootheril $et$ $al$. demonstrated that sulfur-containing heteroatoms enhance ICD efficiency in thiophene dimers, providing a route to control low-energy electron emission and elucidate its role in radiation-induced molecular damage \cite{Mootheril2025}.

ETMD represents another important nonlocal relaxation mechanism that frequently competes with ICD in extended systems \cite{zobeley2001, Sakai2011}. ETMD is initiated by the creation of a core or valence hole that is filled via electron transfer from a neighboring atom or molecule. The released energy subsequently ionizes either the donor [ETMD(2)] or another nearby species [ETMD(3)], defining two distinct ETMD pathways \cite{Jahnke2020}. Recent studies have demonstrated ETMD’s contribution to radiation-induced damage in solvated systems, such as aqueous Al-ion complexes, where multiple water ionizations have been observed \cite{Gopakumar2023}.

Pyrrole, often considered a model heterocycle for nucleobases, constitutes a key structural element of many biologically important molecules, including vitamin B$_{12}$, bile pigments such as bilirubin, the porphyrin ring of heme, and chlorophyll \cite{Andrew2010, Ashfold2006}. Interactions between pyrrole and water closely mimic those in pyrrole-based biomolecular environments. Recent experiments \cite{Zhou2025} revealed a double-ionization–induced decay process in hydrated pyrrole, accompanied by proton transfer that produces deprotonated C$_4$H$_4$N$^+$ and H$_3$O$^+$ cations. Upon inner-valence ionization, isolated pyrrole predominantly undergoes Auger decay, whereas hydration introduces additional relaxation pathways such as ICD and ETMD, depending on the initial vacancy location \cite{skitnevskaya2023, Kumar2025}. These findings underscore the crucial role of the local hydration environment in modulating the competition among decay channels, offering microscopic insight into radiation-induced ultrafast processes in biological systems \cite{boudaiffa2000, skitnevskaya2023, gao2025, Zhang2025}.

To elucidate these mechanisms from first principles, we employ a real-time time-dependent density functional theory (TDDFT) framework that treats electrons and nuclei on an equal footing \cite{Taylor2025, Covington2017, Bubin2012, wang2025}. 
This approach allows us to directly characterize the dynamics of ICD by monitoring ionized-electron emission and monomer charge distribution. Key features such as charge transfer and nonadiabatic effects are explicitly captured, with proton motion identified as a potential gating mechanism \cite{Sharma2020}. The methodology further enables the study of related ultrafast decay channels, including Auger and ETMD processes.
In this work, we apply real-space, real-time TDDFT
\cite{wang2025} combined with Ehrenfest dynamics 
to investigate the ultrafast relaxation dynamics of 
inner-valence–ionized hydrated pyrrole under different initial
vacancy configurations. Our simulations provide microscopic 
insights into the competition between Auger, ICD, and ETMD energy-transfer pathways.

%==============================================
\section{COMPUTATIONAL METHOD}\label{section1}
%==============================================
The molecular dynamics in each set of simulations are modeled by 
real-time TDDFT  on a real-space grid. The Kohn–Sham Hamiltonian
has the following form:
\begin{align}
\hat{H}_{\mathrm{KS}}(t)=&-\frac{\hbar^{2}}{2m}\nabla^{2}+V_{\mathrm{ion}}(\mathbf{r},t)+V_{H}[\rho](\mathbf{r},t)\notag\\&+V_{XC}[\rho](\mathbf{r},t).
\end{align}
Here, $\rho$ is the electron density which is defined as the density
sum over all occupied orbitals
\begin{equation}
\rho(\mathbf{r}, t) = \sum_{k=1}^{N} 2|\psi_k(\mathbf{r}, t)|^2,
\end{equation}
the coefficient 2 accounts for the presence of two electrons in each orbital due to electron spin degeneracy, while k denotes the quantum number for each orbital.
The $V_{ion}$ in Equation (1) represents the external potential induced by ions, expressed using the norm-conserving pseudopotential centered on each ion as provided by Troulier and Martins \cite{Troullier1991}.
$V_H$ denotes the Hartree potential, defined as
\begin{equation}
V_H(\mathbf{r}, t) = \int \frac{\rho(\mathbf{r}', t)}{|\mathbf{r} - \mathbf{r}'|} \, d\mathbf{r}',
\end{equation}
representing the average electrostatic interaction resulting from electron repulsion.
The last term, $V_{XC}$, represents the exchange–correlation potential, which is approximated using the generalized gradient approximation (GGA) developed by Perdew $et$ $al$. \cite{perdew1992}.
Prior to the time-dependent simulations, ground-state DFT calculations are performed to obtain the equilibrium electronic structure of the system, including the electron density, self-consistent Kohn–Sham orbitals, and total energy.
With these quantities established as initial conditions, the time-dependent Kohn–Sham (TDKS) equations are then propagated in real time to evolve the Kohn–Sham orbitals 
$\psi_{k}(\mathbf{r},t)$  according to
\begin{equation}
i\hbar\frac{\partial\psi_k(\mathbf{r},t)}{\partial t}=\hat{H}_{\mathrm{KS}}(t)\psi_k(\mathbf{r},t),
\end{equation}
the equation is solved by time propagation,
\begin{equation}
\psi_k(r,\mathrm{t}+\delta\mathrm{t})\approx\exp\left[-\frac{i\hat{H}_{\mathrm{KS}}(\mathrm{t})\delta\mathrm{t}}{\hbar}\right].
\label{equation5}
\end{equation}
The equation (\ref{equation5}) is approximated by a fourth-order Taylor expansion as follows:
\begin{equation}
\psi_{k}(\mathbf{r},t+\delta t)\approx\sum_{n=0}^{4}\frac{1}{n!}\left(\frac{-i\delta t}{\hbar}\hat{H}_{\mathrm{KS}}(t)\right)^{n}\psi_{k}(\mathbf{r},t).
\end{equation}

The time-propagation operator is applied iteratively for 
$N$ time steps until the final simulation time  $t_{final}$ = $N\cdot\delta t$ is obtained. 
A time step of $\delta t$ = 1 attosecond is employed in all simulations.
In our RT-TDDFT calculations, the Kohn–Sham orbitals are represented on a uniform real-space grid. The accuracy of the simulations depends 
sensitively on the grid spacing, which serves as a key numerical parameter controlling spatial resolution. In the present work, we used a grid spacing of 0.22 \AA, with 139, 122, and 130 grid points along the 
x, y, and z directions, respectively.
The Kohn–Sham orbitals are set to zero at the boundaries of the simulation box. To suppress nonphysical reflections of the electronic density as ionized fragments approach the edges, a complex absorbing potential (CAP) is applied in the outer regions of the box. The CAP follows the form proposed by Manolopoulos \cite{Manolopoulos2002}:

\begin{equation}
-iw(x) = -\frac{i\hbar^{2}}{2m}\left(\frac{2\pi}{\Delta x}\right)^{2}f(y),\quad y = \frac{\left(x-x_{1}\right)}{\Delta{x}},
\end{equation}
where $x_1$ is the starting point of the absorption region, $x_2$ is the end point, $\Delta x = x_2 - x_1$, c = 2.62 is a numerical constant, M is the electron mass, and
\begin{equation}
f(y)=\frac{4}{c^{2}}\left(\frac{1}{(1+y)^{2}}+\frac{1}{(1-y)^{2}}-2\right).
\end{equation}

When the molecule is ionized, the electron density propagates toward the boundaries of the simulation box, where it is absorbed by the complex absorbing potential (CAP). The total number of electrons within the simulation box at time t is given by
\begin{equation}
N(t)=\int_{V}\rho(\mathbf{r},\mathrm{t})d^{3}x,
\end{equation}
where $V$ denotes the volume of the simulation box. The total ionization of the system is then evaluated as $N (0)$ - $N(t)$.
Nuclear motion is treated using the Ehrenfest method, in which nuclei evolve classically under the influence of time-dependent quantum forces. The force acting on the $i$-th nucleus is obtained as the gradient of the total energy with respect to the nuclear coordinate, expressed as
\begin{align}
M_i\frac{d^2\mathbf{R}_i}{dt^2}
=&\sum_{j\neq i}^{N_{\mathrm{ions}}}
\frac{Z_iZ_j(\mathbf{R}_i-\mathbf{R}_j)}{|\mathbf{R}_i-\mathbf{R}_j|^3}
\notag\\
&-\nabla_{\mathbf{R}_{i}}\int
V_{\mathrm{ion}}(\mathbf{r},\mathbf{R}_{i})\rho(\mathbf{r},t)\,d\mathbf{r},
\end{align}
where $M_i$, $Z_i$, and $\mathbf{R}_{i}$ are the mass, pseudocharge
(valence), and position of the ith ion, respectively, and $N_{ions}$ is the total number of ions. The differential equation propagates in time at each time step $\delta t$ in combination with the Verlet algorithm.
\begin{figure}[htbp]
  \centering
  \includegraphics[width=\linewidth]{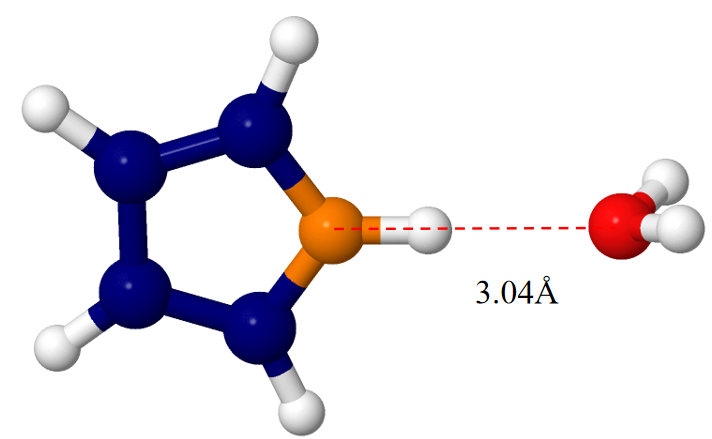}
  \caption{The geometric structure of the hydrated pyrrole dimer used in this study. The nuclear distance between O and N atoms is in angstroms.}
  \label{fig:1}
\end{figure}

Fig.~\ref{fig:1} illustrates the optimized geometry of the hydrated pyrrole dimer, obtained using the B3LYP/aug-cc-pVTZ method. The distance between the oxygen and nitrogen atoms is 3.04~\AA. In this configuration, the water molecule acts as an electron-density donor, while pyrrole serves as the electron-density acceptor. As discussed below, we investigate the electron-relaxation dynamics following inner-valence ionization of the hydrated pyrrole dimer.
To represent the initial ionized state, a ground-state DFT calculation is first performed to obtain the electronic structure of the neutral system. An initial vacancy is then introduced by removing an electron from a specified molecular orbital, after which the system is propagated in real time using TDDFT.
In the present calculations, a complex absorbing potential (CAP) is employed to effectively remove ionized electrons from the simulation domain. The charge of each monomer is evaluated by integrating the time-dependent electron density over regions centered on each molecular subunit.

%==============================================

%=================================================
\section{RESULTS AND DISCUSSION}\label{section2}
%=================================================
In this study, we investigate the possible decay channels that arise following inner-valence ionization of either the water molecule or the heteroatom in the pyrrole–H$_2$O complex. Multiple dynamical trajectory simulations were performed, with the initial atomic velocities randomly sampled from a Boltzmann distribution at 300~K, a standard procedure for initializing nuclear motion \cite{wang2025, Alonso2008, Taylor2025}.
A total of 90 trajectories were propagated, with the number constrained by computational cost. When the initial hole for O 2s$^{-1}$ state was created on the water molecule, 30 trajectories were simulated, of which 21 led to ICD and 9 to ETMD. In contrast, when the hole for N 2s$^{-1}$ state was created on the pyrrole molecule, 60 trajectories were performed, yielding 30 ICD and 30 Auger decay events.
To elucidate the underlying relaxation mechanisms, several representative trajectories are analyzed in detail below.

\begin{figure}[htbp]
  \centering
  \includegraphics[width=\linewidth]{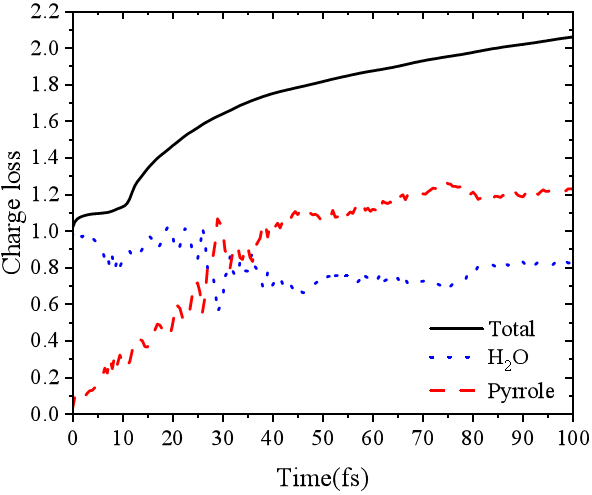}
  \caption{Time-dependent charge loss of pyrrole-water with the initial vacancy on water during the ICD process.}
  \label{fig:2}
\end{figure}

Fig.~\ref{fig:2} presents the time-dependent charge loss of the hydrated pyrrole dimer during the ICD process when the initial vacancy is created on the water molecule. As shown, the system begins to decay following ionization of the oxygen 2s electron in water. At the initial moment, the water molecule is cationic, while pyrrole remains neutral. Over the entire 100~fs time window, the charge loss on water remains nearly constant at approximately 0.8. In contrast, the charge loss on pyrrole increases sharply at around 5~fs, reaching a maximum of about 1.1 at 29~fs, and subsequently fluctuates between 0.9 and 1.2 up to 100~fs. 
These fluctuations are due to minor oscillations in the electron density, such fluctuations can seen in all of the charge loss plots.
These results indicate that most of the net charge loss occurs on pyrrole, whereas water maintains its ionic character throughout the simulation. This behavior is consistent with the characteristic features of the ICD mechanism, confirming that the present simulation successfully captures the ICD process in the hydrated pyrrole dimer. The maximum total charge loss is approximately 2.1, slightly higher than the theoretical expectation of 2; however, such deviation is reasonable considering the intrinsic limitations of TDDFT and the spatially defined complex absorbing potential (CAP). The gradual increase in total charge loss arises from the finite time required for the ionized wavefunction to reach the CAP region, as discussed in our previous work \cite{wang2025}. Notably, a pronounced rise in total charge loss is observed between 15 and 40~fs, suggesting that Coulombic decay predominantly occurs within this time interval.

%=================================================
\begin{figure}[htbp]
  \centering
  \includegraphics[width=\linewidth]{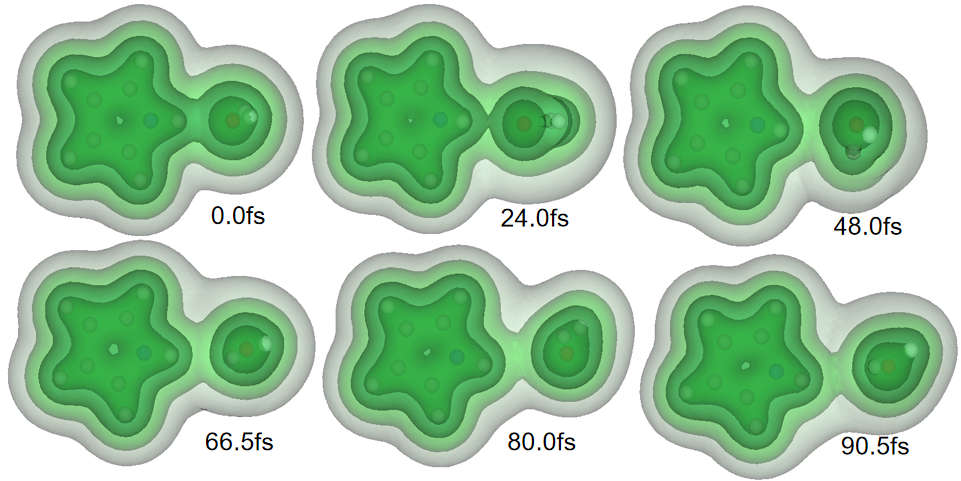}
  \caption{Time-dependent electron density snapshot of pyrrole-water with the initial vacancy on water during the ICD process.}
  \label{fig:3}
\end{figure}

The time-dependent electron density provides further insight into the ICD dynamics of the pyrrole–water system. Fig.~\ref{fig:3} displays representative snapshots of the evolving electron density. Between 0 and 24~fs, the two hydrogen atoms of the water molecule elongate synchronously along the direction opposite to the N–H···O hydrogen bond, reaching their maximum extension at 24~fs. Thereafter, they undergo synchronous in-plane oscillations parallel to the pyrrole ring. Concurrently, as the ICD process proceeds, both water and pyrrole molecules become positively charged, resulting in pronounced Coulombic repulsion and a gradual increase in their intermolecular separation.

%=================================================
\begin{figure}[htbp]
  \centering
  \includegraphics[width=\linewidth]{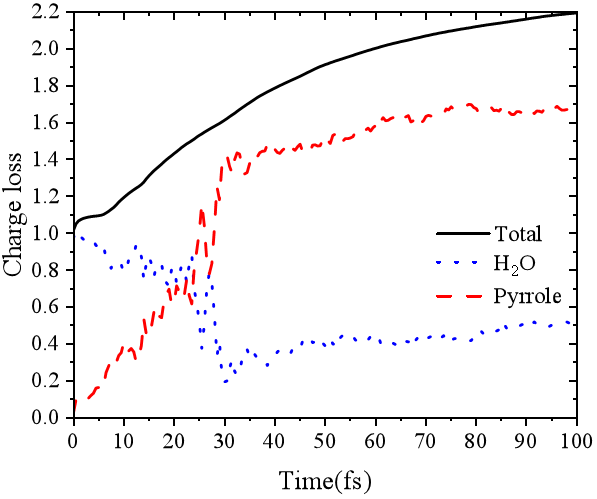}
  \caption{Time-dependent charge loss of pyrrole-water with the initial vacancy on water during the ETMD process.}
  \label{fig:4}
\end{figure}

%=================================================
\begin{figure}[htbp]
  \centering
  \includegraphics[width=\linewidth]{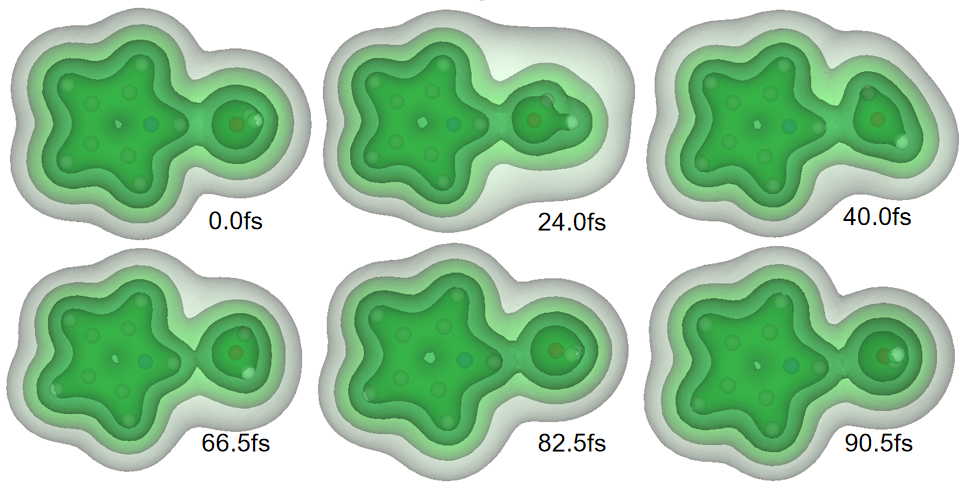}
  \caption{Time-dependent electron density snapshot of pyrrole-water with the initial vacancy on water during the ETMD process.}
  \label{fig:5}
\end{figure}

The ionization of the oxygen 2s electron in water initiates not only the ICD process discussed above but also an alternative relaxation pathway—ETMD.
Fig.~\ref{fig:4} presents the time-dependent charge loss of the hydrated pyrrole dimer during this process. Between 0 and 25~fs, the charge loss on the water molecule gradually decreases from 1.0 to 0.2. After 30~fs, it changes only slightly, reaching a final value of 0.5 at 100~fs. In contrast, the charge loss on pyrrole rises from 0 at the initial moment to 1.5 at 30~fs, followed by a slower increase to 1.7 at 100~fs. At the end of the simulation, the water molecule is nearly neutral, while the pyrrole molecule has lost approximately two electrons, consistent with the ETMD mechanism.
A pronounced charge transfer between the two monomers occurs around 25–30~fs, indicating that the ETMD process likely takes place within this time window. Although previous studies have reported that ETMD generally proceeds more slowly than ICD \cite{skitnevskaya2023}, our simulations reveal that both processes can occur on comparable femtosecond timescales.
The corresponding time-resolved electron density snapshots are shown in Fig.~\ref{fig:5}. The variations in electron density mainly originate from the stretching vibrations of the two hydrogen atoms in the water molecule. Since the water molecule becomes nearly neutral by the end of the process, the intermolecular distance remains almost unchanged at 100~fs.

Skitnevskaya $et$ $al$. \cite{skitnevskaya2023} investigated the relaxation dynamics of inner-valence vacancies in hydrated pyrrole by computing single- and double-ionization spectra using the third-order algebraic diagrammatic construction (ADC(3)) and second-order ADC(2)) methods, respectively. Their results showed that, following inner-valence ionization of the water molecule, only nonlocal relaxation channels—ICD and ETMD—are energetically accessible. This finding is fully consistent with the conclusions of the present study.

Kumar $et$ $al$. \cite{Kumar2025} examined the energetics of electron decay in inner-valence–ionized pyrrole–water pairs (C\textsubscript{4}H\textsubscript{5}N–H\textsubscript{2}O) using the complete active space second-order perturbation theory (CASPT2) method to determine ionization potentials (IP) and double-ionization potentials (DIP). For the O-2s vacancy, their calculations predicted that both ICD and ETMD channels are energetically feasible in addition to Auger decay. In contrast, our time-dependent simulations did not reveal any Auger decay events. This discrepancy likely originates from the fact that the CASPT2 calculations were performed at fixed nuclear geometries, where the neglect of nuclear relaxation effects may influence the relative energetics of competing decay pathways.

%=================================================
\begin{figure}[htbp]
  \centering
  \includegraphics[width=\linewidth]{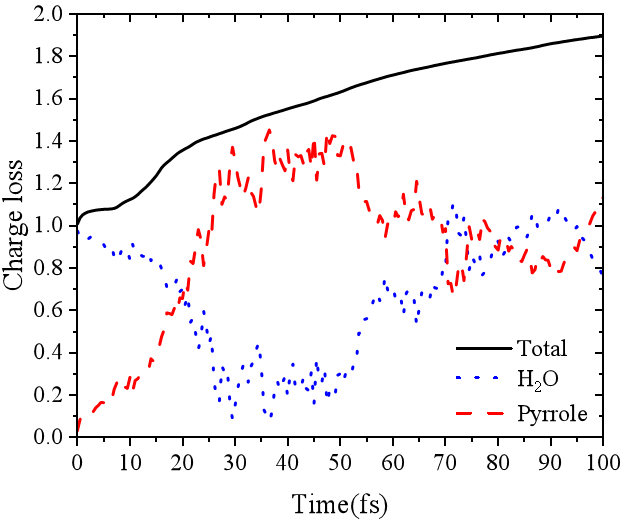}
  \caption{Time-dependent charge loss of pyrrole-water with the initial vacancy on water during the other ICD process.}
  \label{fig:55}
\end{figure}
%=================================================
\begin{figure}[htbp]
  \centering
  \includegraphics[width=\linewidth]{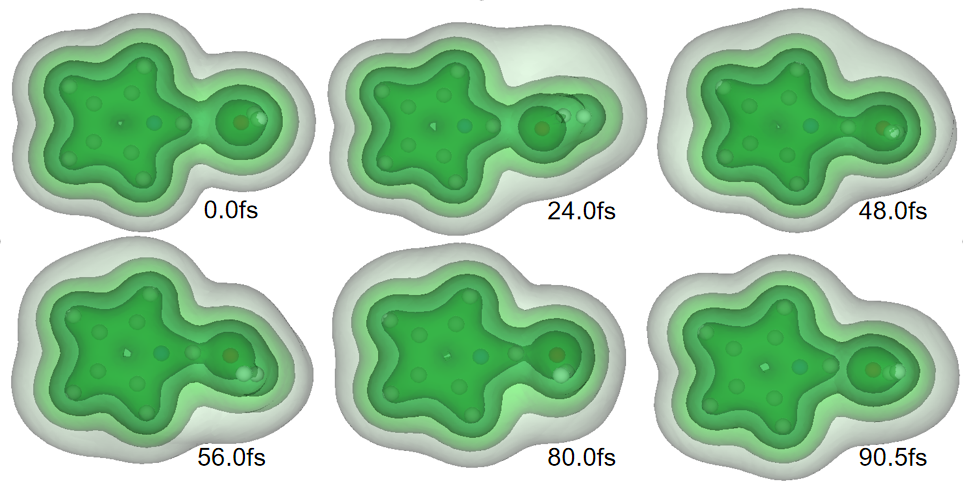}
  \caption{Time-dependent electron density snapshot of pyrrole-water with the initial vacancy on water during the other ICD process.}
  \label{fig:66}
\end{figure}

In hydrated biomolecular clusters, inner-valence ionization of water molecules can initiate ICD, as previously confirmed in hydrated tetrahydrofuran clusters \cite{Ren2018}, consistent with the present findings. In addition to ICD, our simulations predict a competing decay channel—ETMD—in which the final state corresponds to a doubly ionized organic molecule. Both ICD and ETMD involve charge transfer and electron emission, and their interplay may lead to mutual conversion between the two processes.
This competition is also observed in our simulations, as illustrated by the time-dependent charge loss shown in Fig.~\ref{fig:55}. At $t=0$~fs, the water molecule loses one electron, while pyrrole remains neutral. Subsequently, charge transfer occurs from pyrrole to water, reducing the charge loss of water to approximately 0.2 and increasing that of pyrrole to about 1.4 within 30–50~fs. This behavior is consistent with the characteristics of ETMD, where the water molecule remains nearly neutral and the pyrrole molecule carries a charge exceeding +1.
The corresponding charge redistribution can also be visualized in the electron density maps at 0~fs and 24~fs, shown in Fig.~\ref{fig:66}. During this period, the total charge loss of the system increases from 1.0 to 1.6, with the additional loss primarily originating from pyrrole. At 100~fs, the charge losses of water and pyrrole reach approximately 0.8 and 1.1, respectively—features that are characteristic of the ICD process, indicating a dynamical transition from ETMD to ICD during the relaxation.

%=================================================
\begin{figure}[htbp]
  \centering
  \includegraphics[width=\linewidth]{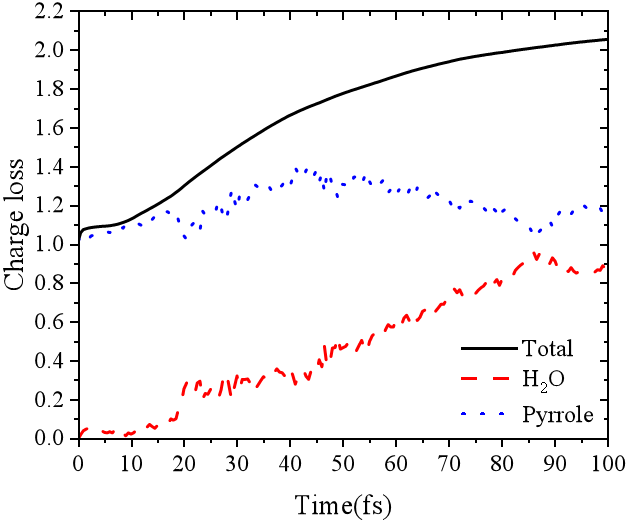}
  \caption{Time-dependent charge loss of pyrrole-water with the initial vacancy on pyrrole during the ICD process.}
  \label{fig:6}
\end{figure}

\begin{figure}[htbp]
  \centering
  \includegraphics[width=\linewidth]{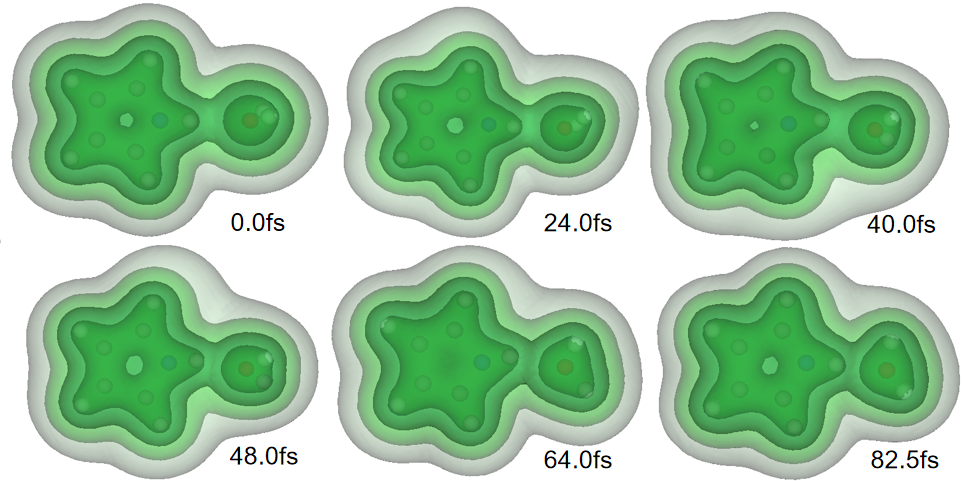}
  \caption{Time-dependent electron density snapshot of pyrrole-water with the initial vacancy on pyrrole during the ICD process.}
  \label{fig:7}
\end{figure}

We also examined the electron decay mechanism when the initial vacancy is localized on the pyrrole molecule. Fig.~\ref{fig:6} shows the time-dependent charge loss of pyrrole and water. The total charge loss gradually increases from 1.0 at $t = 0$~fs to 2.1 at 100~fs. Within the first 15~fs, the charge loss on the water molecule remains negligible, then rises sharply to about 0.3 at 20~fs, and continues to increase, reaching approximately 0.9 at 100~fs. In contrast, the charge loss on pyrrole fluctuates between 0.9 and 1.3 throughout the same period, with a final value of 1.2 at 100~fs. Since the dominant charge loss occurs on the water molecule, this behavior is characteristic of the ICD mechanism. Although the gradual increase in total charge loss makes it difficult to pinpoint the precise onset time of ICD, the results suggest that the process occurs within a timescale of roughly 100~fs.
Fig.~\ref{fig:7} presents selected snapshots of the electron density during the ICD process. At the initial time, the oxygen atom of the water molecule lies nearly in the plane of the pyrrole ring; by 100~fs, it deviates from this plane by approximately 35°. During this period, the pyrrole ring undergoes a transition from a planar to a slightly twisted geometry, reflecting structural relaxation that minimizes electrostatic repulsion between the two positively charged fragments.

%=================================================

%=================================================
\begin{figure}[htbp]
  \centering
  \includegraphics[width=\linewidth]{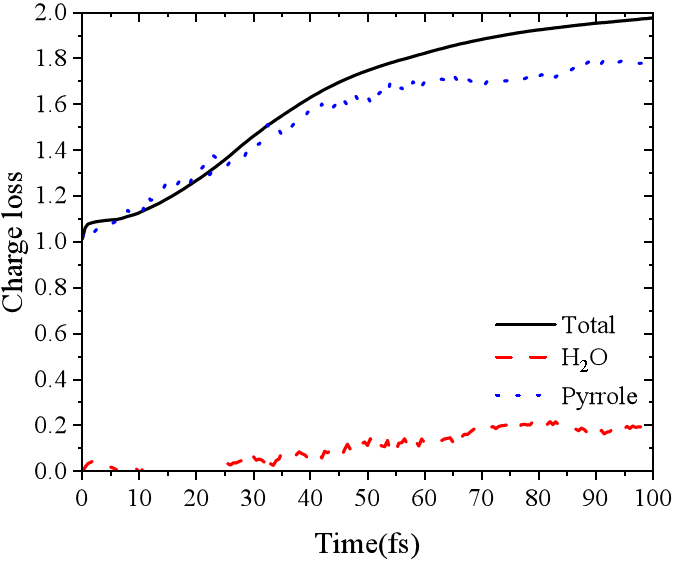}
  \caption{Time-dependent charge loss of pyrrole-water with the initial vacancy on pyrrole during the Auger process.}
  \label{fig:8}
\end{figure}

Finally, we examine the cases in which Auger decay occurs following inner-valence ionization of the pyrrole molecule. The time-dependent charge loss is shown in Fig.~\ref{fig:8}. During the first 30~fs, the total charge loss originates primarily from pyrrole, while the charge loss on water remains negligible. As the system evolves, a small but gradual increase in charge loss on water is observed, reaching less than 0.2 within the simulated time window. After 30~fs, the total charge loss continues to rise, eventually reaching a value of approximately 2.0 at 100~fs, with pyrrole contributing about 1.8.
This behavior indicates that the majority of electron emission occurs from pyrrole, consistent with the characteristics of Auger decay. The minor charge redistribution toward water likely reflects weak intermolecular coupling during the relaxation. These results thus confirm that the inner-valence–ionized pyrrole primarily relaxes via Auger decay under the present conditions.
%==============================================
\begin{figure}[htbp]
  \centering
  \includegraphics[width=\linewidth]{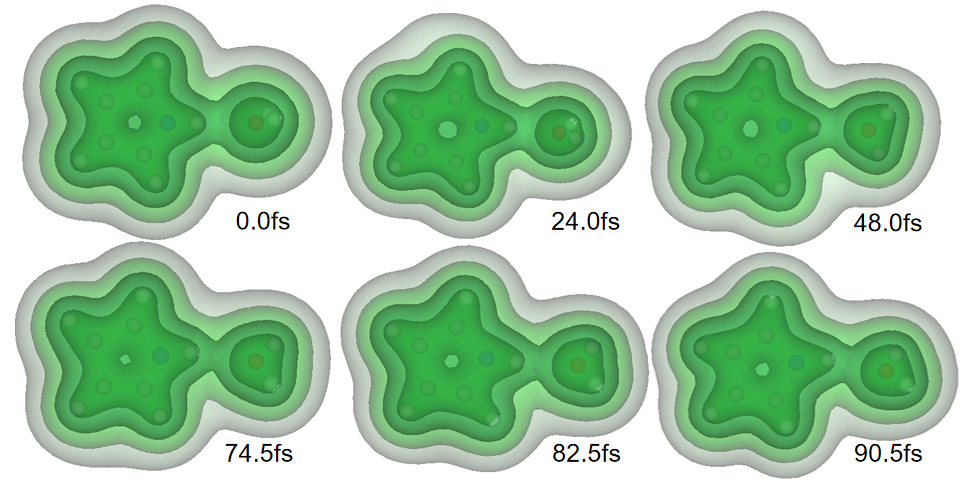}
  \caption{Time-dependent electron density snapshot of pyrrole-water with the initial vacancy on pyrrole during the Auger process.}
  \label{fig:9}
\end{figure}

Fig.~\ref{fig:9} presents time-resolved electron density snapshots for the Auger decay process following inner-valence ionization of pyrrole in the hydrated pyrrole dimer. Ionization of a valence electron within pyrrole initiates Auger decay, producing a dicationic species (Pyrrole$^{2+}$) and inducing a slight distortion of the pyrrole ring from its planar geometry. The resulting redistribution of charge within pyrrole drives O–H stretching and torsional vibrations in the surrounding water molecules.
Because the energy released during Auger decay is largely localized within the pyrrole moiety, the water molecules primarily absorb vibrational energy through hydrogen bonding and electrostatic coupling. As shown in Fig.~\ref{fig:8}, the water molecules remain nearly neutral throughout the 0–100~fs interval, exhibiting only minor charge redistribution associated with the coupling between nuclear motion and electronic dynamics rather than genuine ionization.

For inner-valence ionization of the heterocycle, the hydrated pyrrole system can relax via either Auger decay or ICD. Our results are consistent with the general conclusions of Skitnevskaya $et$ $al$.~\cite{skitnevskaya2023} and Kumar $et$ $al$.~\cite{Kumar2025}. However, Skitnevskaya $et$ $al$. suggested that Auger decay is the dominant relaxation pathway, independent of whether the initial hole is located on the water molecule or on the pyrrole ring. In contrast, our real-time simulations indicate that when the vacancy is on the water molecule, Auger decay is suppressed and ICD becomes the primary relaxation channel. Even when the hole is localized on pyrrole, the occurrence probability of Auger decay does not exceed about 50\%.

Previous studies on hydrated clusters have demonstrated that proton transfer generally occurs when the initial hole is located on the proton-donating site of a hydrogen bond \cite{wang2025,Ren2018}. Kumar $et$ $al$.~\cite{Kumar2025} investigated the relaxation of a hole localized on the nitrogen atom in hydrated pyrrole. By extending the N–H bond and evaluating the ionization potentials of the neutral molecule using the CASPT2 method, they proposed that proton transfer may take place. In contrast, no proton transfer events were observed in our simulations across 60 trajectories, suggesting that the probability of such a process is relatively low. Nonetheless, the possibility of proton transfer cannot be completely excluded.

%==============================================
\section{SUMMARY}\label{section3}
%==============================================
We have investigated the ultrafast relaxation dynamics following inner-valence ionization in the hydrated pyrrole dimer by employing real-space real-time time-dependent density functional theory (RT-TDDFT) combined with Ehrenfest dynamics. This approach enables a unified treatment of coupled electronic and nuclear motion, thereby capturing charge transfer, ionization, and structural evolution on the femtosecond timescale. The simulations, carried out within the GGA for the exchange–correlation functional, reveal a pronounced dependence of the decay pathways on the location of the initial vacancy.

When the initial vacancy is created on the water molecule, the system predominantly undergoes ICD with a probability of 0.63. This process involves energy transfer to pyrrole, followed by secondary ionization and eventual Coulomb-driven dissociation. The remaining trajectories (probability 0.37) proceed via ETMD, characterized by charge redistribution that drives pyrrole toward a dicationic state while partially neutralizing the water molecule. Both processes occur within approximately 100 fs, with ICD exhibiting a distinct onset between 15 and 40 fs, consistent with previous observations in hydrated clusters \cite{wang2025,wang2024}. Notably, charge transfer from pyrrole to water observed in our simulations demonstrates the dynamical evolution from ETMD to ICD, highlighting the close competition between these two nonlocal decay channels.

In contrast, ionization localized on the pyrrole moiety results in comparable probabilities ($\approx$ 0.5) for ICD and Auger decay. In the ICD channel, energy is transferred to the water molecule, inducing vibrational and rotational excitations without immediate fragmentation. In the Auger process, relaxation remains confined to the pyrrole ring, yielding a dicationic species accompanied by slight geometric distortion. The localized nature of Auger decay is evident from the minimal charge redistribution toward the water molecule, distinguishing it from the strongly delocalized character of ICD.

The pronounced sensitivity of these decay pathways to the initial vacancy location underscores their potential relevance to radiation-induced damage in aqueous and biological environments. The prevalence of ICD and ETMD implies enhanced production of secondary low-energy electrons, which are known to play a crucial role in radiation chemistry and genotoxic processes. Overall, our results provide a microscopic framework for understanding intermolecular energy-transfer dynamics in hydrated molecular systems, with implications for photostability and radiation therapy applications.

Our simulations utilized the mean-field Ehrenfest method, which
computes nuclear dynamics on an averaged energy landscape weighted by
electronic state populations. Other available approaches include
trajectory surface hopping (TSH), where numerous independent classical
paths model nuclear wave packet dynamics on separate Born-Oppenheimer
potential surfaces, and the multiple spawning (MS) method, which
expresses the nuclear wave function through Gaussian basis sets that
evolve along classical trajectories.
These three methods share a common feature: nuclear motion is
described classically—although MS uses these classical paths as an
auxiliary construct supporting quantum nuclear dynamics. As a result,
all methods necessitate computing electronic characteristics
(energies, energy gradients, coupling elements, etc.) at the
instantaneous classical nuclear configurations throughout the
simulation. Ref. \cite{doi:10.1021/acs.chemrev.7b00577} provides a detailed discussion of the
advantages and drawbacks of these techniques.
We plan to extend our computational framework to include TSH
methodology and compare the relative effectiveness of Ehrenfest and
TSH approaches.

Another possibility is to maintain the Ehrenfest formalism while
improving its precision. A core weakness of real-time TDDFT stems from
the need to evaluate exchange-correlation functionals for strongly
perturbed, non-equilibrium configurations far removed from
ground-state conditions, despite these approximate functionals being
calibrated using ground-state exchange-correlation data. This
challenge can be overcome by recasting the Ehrenfest dynamics within a
many-body basis set formulation that incorporates TDDFT-generated
information—a strategy that has demonstrated improved accuracy in
previous studies \cite{doi:10.1021/acs.jpclett.1c02020}. We also plan
to extend our studies to incorporate this approach.
%==============================================
\begin{acknowledgments}
%==============================================
This work was supported by the Natural Science Foundation of Henan Province under Grant No. 252300421490 and by
the National Science Foundation (NSF) under Grant No. DMR-2217759. 

 %==============================================
\end{acknowledgments}
%==============================================
\nocite{*}
\bibliography{ref.bib} %Produces the bibliography via BibTeX.
%==============================================
\end{document}